\documentclass{moriond}

\def\Journal#1#2#3#4{{#1} {\bf #2}, #3 (#4)}

\def\PLB{{\em Phys. Lett.}  B}

\def\PRD{{\em Phys. Rev.} D}

\def\be{\begin{equation}}
\def\ee{\end{equation}}
\def\bea{\begin{eqnarray}}
\def\eea{\end{eqnarray}}

\newcommand{\Photo}{\includegraphics[height=35mm]{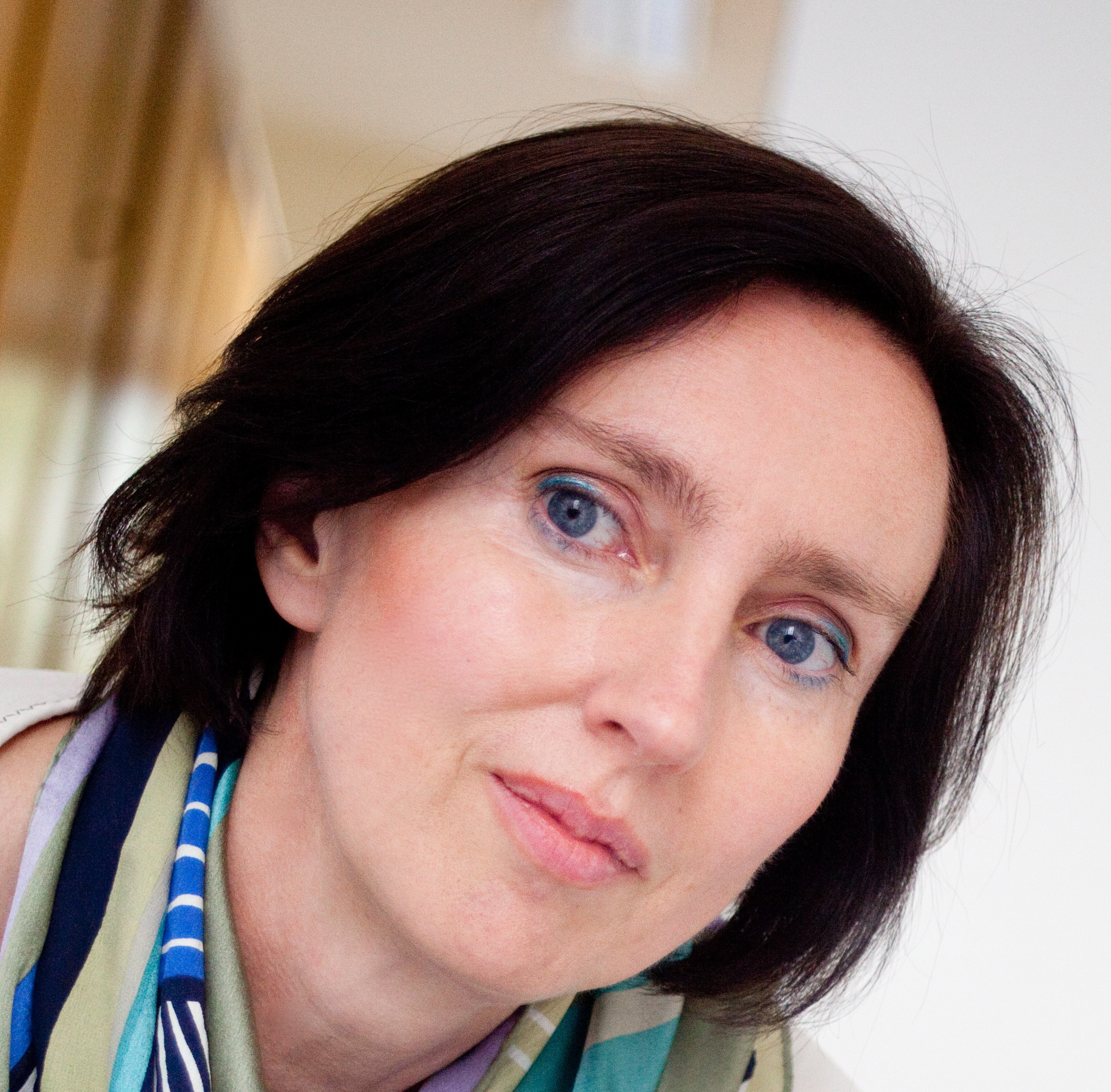}}

\begin{document}
\vspace*{4cm}
\title{TECHNIQUES AND RESULTS OF NEUTRAL LONG-LIVED PARTICLE SEARCHES IN ATLAS AND CMS IN LHC RUN 2}

\author{ CLAUDIA-ELISABETH WULZ, for the ATLAS and CMS Collaborations }

\address{Institute of High Energy Physics of the Austrian Academy of Sciences,\\
Nikolsdorfergasse 18, 1050 Vienna, Austria}

\maketitle\abstracts{
Selected results of searches for neutral long-lived particles with displaced and delayed jets, trackless jets and emerging jets, as well as displaced muons, obtained with data recorded at a centre-of-mass energy of 13 TeV by the ATLAS and CMS Collaborations during LHC's Run 2 are presented, as well as challenges and techniques.
}

\footnotetext{Copyright 2019 CERN for the benefit of the ATLAS and CMS Collaborations. CC-BY-4.0 license.}

\section{Introduction}
The standard model (SM) of particle physics includes a number of uncharged long-lived particles (LLP) such as neutrinos, neutrons or neutral B, D, and K mesons. Exotic, but so far undetected long-lived particles emerge in many theories extending the standard model. We focus on detecting such neutral particles and present selected recent results of searches by the ATLAS and CMS collaborations \cite{ATLAS,CMS} at the LHC in proton-proton collisions at a centre-of-mass energy of $\sqrt s = 13$~TeV. 
In the context of these searches LLP's are defined as particles that do not decay instantly at the interaction point, where they are produced, but at some measurable distance within the detector. They may also remain stable within the detector volume. 

A plethora of models predict long-lived particles. Supersymmetry (SUSY) in various incarnations, dark matter scenarios, or portal interactions between a hidden sector and the SM are sources of LLP's. Among the SUSY models are R-parity violating SUSY,  gauge- and anomaly-mediated SUSY breaking scenarios, split SUSY, or stealth SUSY. In gauge-mediated scenarios the gravitino is the lightest supersymmetric particle (LSP). 
The striking feature of split supersymmetry is that the gluino is a quasi-stable particle with a lifetime that could be up to a hundred seconds long. Scalars except the Higgs boson are ultra-heavy. In stealth SUSY R-parity is preserved, but missing energy signatures are lacking. This becomes possible through a new hidden sector of particles at the weak scale. Since SUSY breaking is small, the fermion and boson pairs will be almost degenerate in mass. 
In hidden valley scenarios particles of the hidden sector such as long-lived dark photons or dark hadrons may decay to standard model particles. Left-right symmetric models predict heavy right-handed neutrinos. Axions or axion-like particles are also expected to have long lifetimes.  

A particle's proper lifetime is inversely proportional to its mass, and proportional to the squared matrix element for the decay into its daughter particles and the phase space for the decay. Heavy particles normally decay quickly into lighter ones, unless their decay is suppressed by some mechanism. Suppression can occur if couplings are small, if the decaying particle and the decay products are almost degenerate in mass, or if the decay is mediated by very heavy virtual exchange particles.

\section{Challenges and techniques}

Searches for particles with proper lifetimes $c\tau > 1$~mm have been ongoing for many years at the ATLAS and CMS experiments, even though the detectors were a priori designed and optimised for promptly produced particles. However, no signals have been observed so far. Therefore preparations for Run 3 of the LHC and the corresponding detector upgrades have a strong focus on LLP's. In parallel with new theoretical developments clever ideas for triggering, data acquisition, event reconstruction and analysis have been and are being worked out, exploiting the detectors in ways they were not necessarily designed for.

\begin{figure} [hbt]
\centerline{\includegraphics[width=0.40\linewidth]{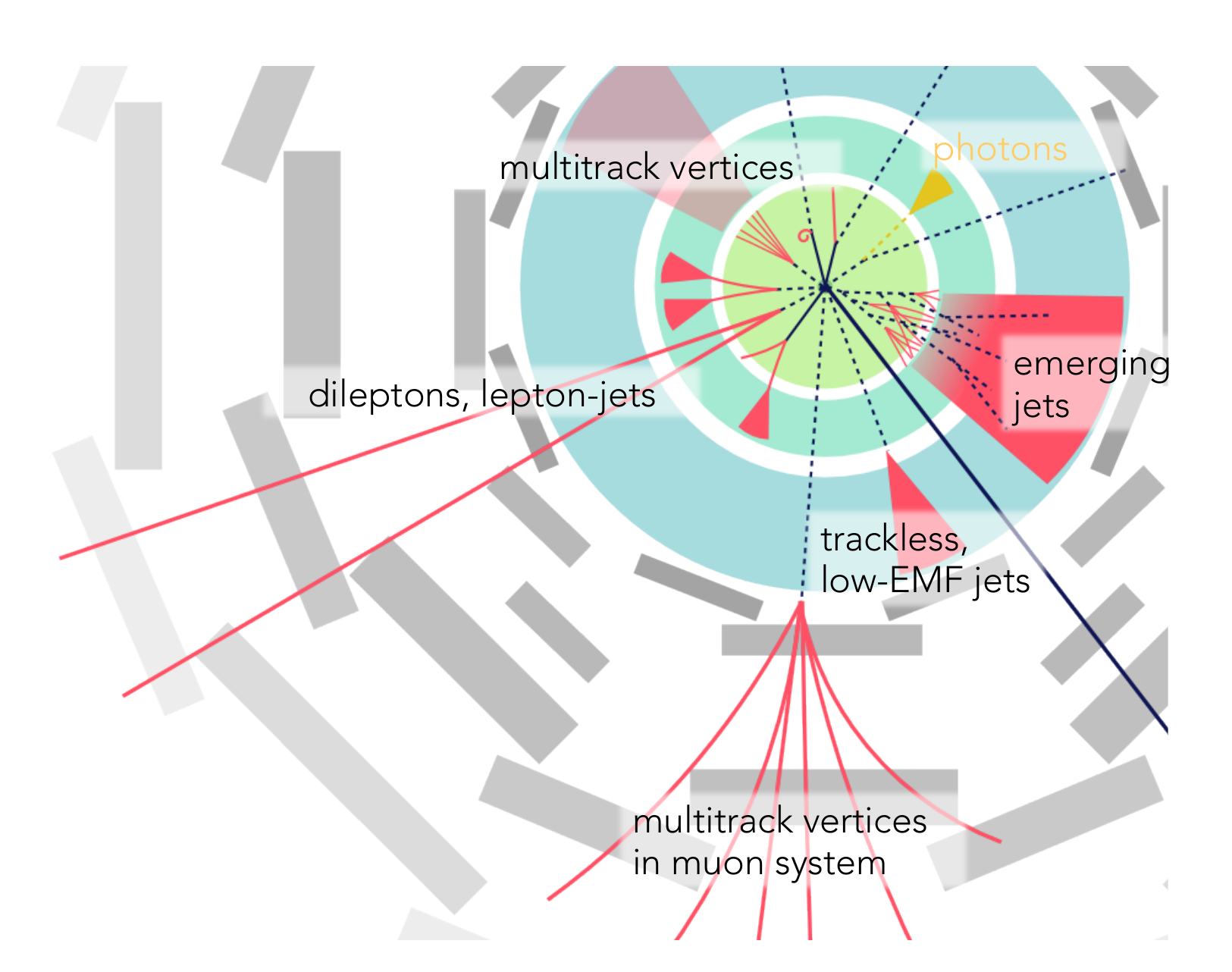}}
\caption[]{Displaced decays in a collider detector (adapted from Heather Russell, with permission).}
\label{fig:Russell}
\end{figure}

Several challenges have to be overcome. The first challenge comes from physics itself. Unusual or unexpected signatures of long-lived particles may arise. For example, decays could occur outside the usual, dedicated detectors. Monte-Carlo generators need to be developed in accord with theoretical progress, and therefore samples with simulated data may not be readily available. 
In order to be as model-independent as possible, searches are usually signature-driven. One looks for displaced or delayed decays, or even particles trapped in detector material. The MoEDAL experiment \cite{Moedal} at CERN, for example, will search for magnetic monopoles in sections of an old CMS beam pipe. 
Displaced decays with respect to the primary interaction vertex can manifest themselves in a number of ways, for instance as displaced multi-track vertices, photons, dileptons or lepton-jets. Displaced jets may have particular properties. Jets with multi-track vertices may appear not only in the inner tracking system, but also in the muon chambers. The characteristic signature of emerging jets \cite{Schwaller} are displaced tracks and many different vertices inside the jet cone. Trackless jets, on the other hand, have a low fraction of electromagnetic energy besides no tracks in the inner tracking detector. Figure \ref{fig:Russell} shows a sketch of the different displaced decays in a collider detector. Another signature for LLP's are delayed decays, which require out-of-time detection.  

Trigger, reconstruction and data analysis pose further challenges. Events with LLP's may be lost due to inadequate triggers or triggers with low efficiency. A possible solution is to relax requirements, taking care not to exceed trigger rate limits. Another method is to trigger on prompt particles produced in association with long-lived ones. For particles with very long lifetimes one can also investigate bunch crossings subsequent to the collision of interest. A few special runs with triggers during gaps in LHC bunch trains have already been taken.
Most subdetectors do not provide timing information, or their time resolution is not good enough. CMS, for example, will therefore implement a novel muon timing detector (MTD) with a time resolution of 30~ps \cite{MTD}. Standard event records will have to be expanded to add timing information, where available. Novel ideas for data acquisition that have already been used by the LHC experiments can particularly benefit the search for LLP's. These are data parking and data scouting, also called trigger-level analysis. Parking means that full raw data are stored, but not immediately processed. Scouting on the other hand means that  only a reduced event information is stored, but at rates that can be much higher than the usual bandwidth allocations would permit. For example, searches for dimuon resonances with scouting data could thus be expanded to displaced dimuon searches.

The standard physics object reconstruction is often inadequate for long-lived particles, and systematic uncertainties, relevant for the data analysis, have to be specially calculated. Initial basic requirements need to be revisited in order to avoid that events get rejected at early stages of the reconstruction. The interaction point constraint in triggering or reconstruction is not usable. Track finding is generally optimised for prompt particles, therefore dedicated tracking algorithms need to be developed. Secondary vertex finding is usually tailored to b hadrons, so the b tagging algorithms have to be extended to work also for impact parameters larger than a few millimetres. The particle flow algorithm \cite{PF} that CMS is using needs to be adapted, since by design it uses and combines all available event information from the subdetectors, under the assumption of promptly produced particles. For all experiments, the reconstruction of non-promptly produced electrons and taus needs particular future development efforts. Instrumental effects, where  possible, should be estimated from data.  

Last but not least, the treatment of backgrounds is very challenging. Standard model particles such as K$\mathrm{_L}$ or b hadrons may contribute. Both in-time and out-of-time pileup have to be considered as well. While in-time pileup -- additional proton-proton collisions occurring in the same bunch crossing as the collision of interest -- is also a background in the detection of prompt particles, out-of-time pileup occurring in adjacent bunch crossings is of particular concern in the search for particles with very long lifetimes. But one can actually also use in-time pileup events to search for LLP's \cite{Nachmann}, for example by studying displaced particles with low transverse momenta, which could arise from $u\bar u d \bar d s \bar s$ sexaquarks \cite{Sexaquarks} scattering in detector material or the beam pipe. While it is practically impossible to trigger on such particles directly, they could be present in pileup interactions that are anyway recorded along with a primary interaction that was triggered by a standard algorithm. The added benefit is that there is nearly no selection bias. 
Cosmic rays are of course a major background in LLP searches, since they can easily lead to false displaced decay vertices. Accelerator backgrounds are also relevant, although they are usually well under control. Examples of these are beam halo remnants or satellite bunches. The first come from protons scattered off LHC collimators upstream of the experiments, and the latter originate from collisions of low-luminosity bunches resulting from the bucket structure generated by the accelerator's radio frequency cavities. Interactions with detector material and electronic noise can also be sources of background. Care must be taken for instance that genuine decays in calorimeters are not mistaken for noise. Special mitigation procedures for anomalous detector signals, for example spike-cleaning algorithms in the electromagnetic calorimeter of CMS, must be checked and adapted in order to avoid rejecting interesting long-lived particle signatures.

\section{Results}
The following section presents selected results on displaced and delayed jets, trackless jets and emerging jets, as well as displaced muons, obtained with data recorded at a centre-of-mass energy of 13 TeV.

\subsection{Displaced jets}\label{subsec:DisplacedJets}
\begin{figure} [hbt]
\begin{minipage}{0.33\linewidth}
\centerline{\includegraphics[width=0.9\linewidth]{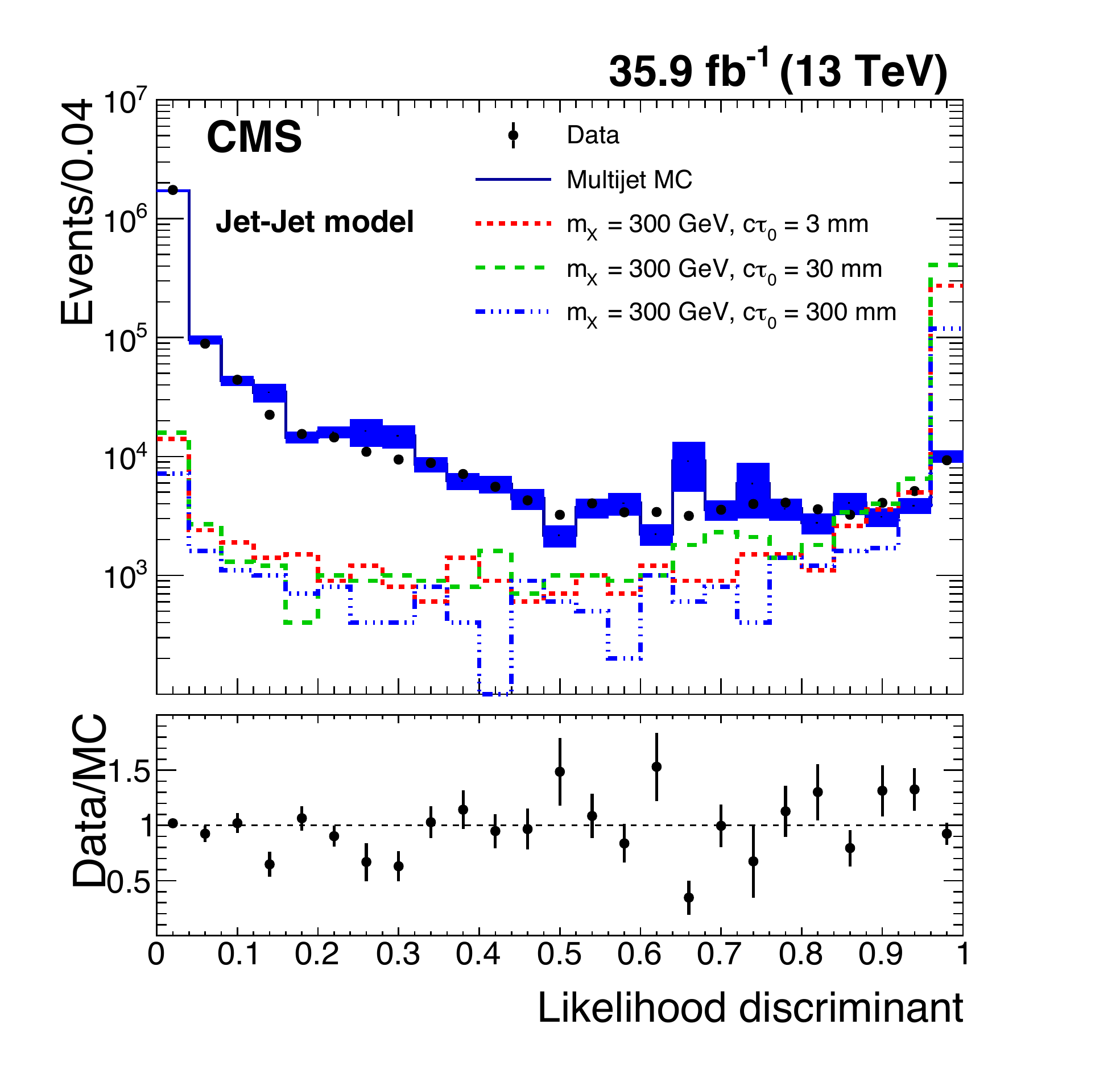}}
\end{minipage}
\hfill
\begin{minipage}{0.32\linewidth}
\centerline{\includegraphics[width=0.9\linewidth]{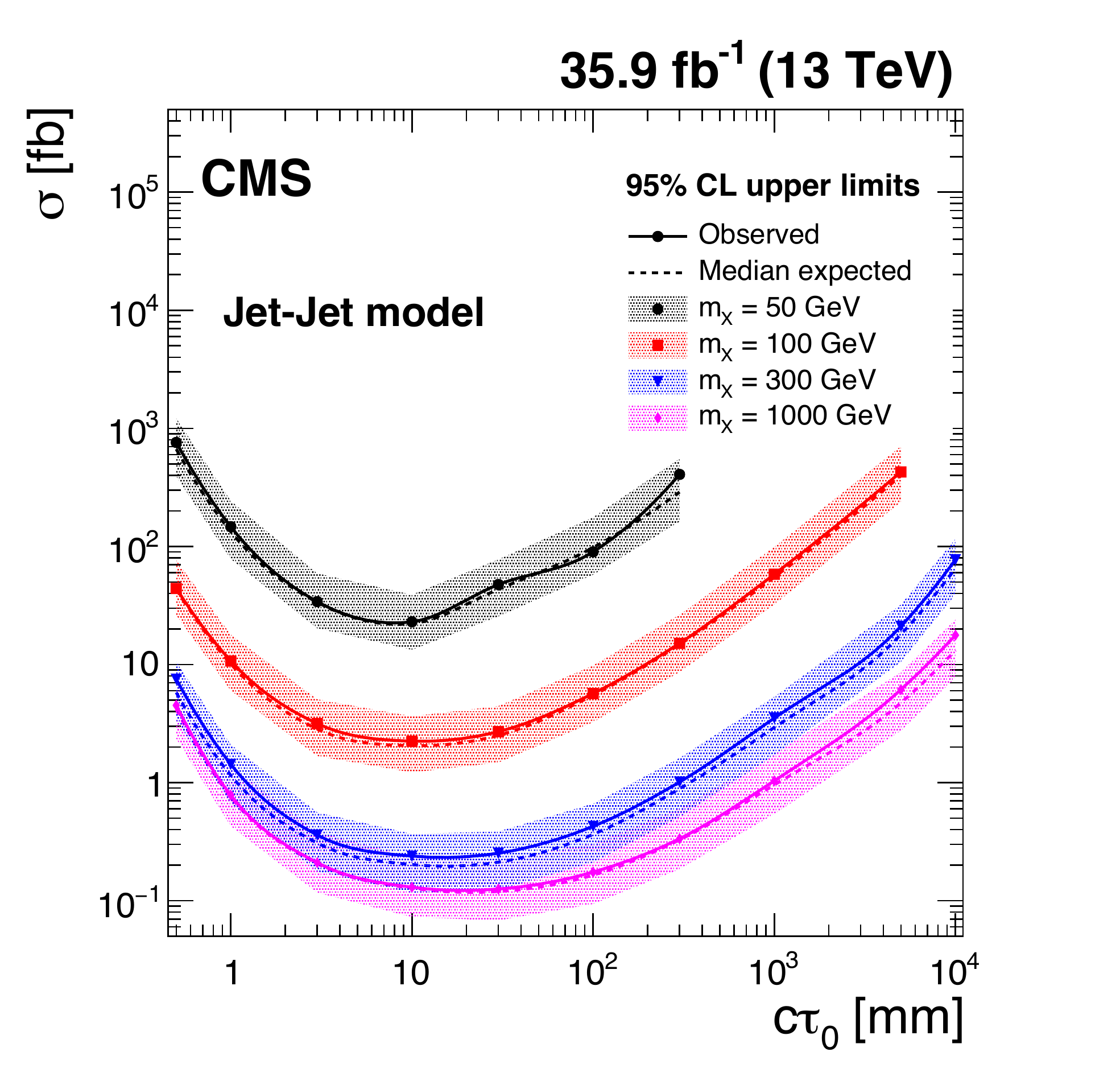}}
\end{minipage}
\hfill
\begin{minipage}{0.32\linewidth}
\centerline{\includegraphics[width=0.9\linewidth]{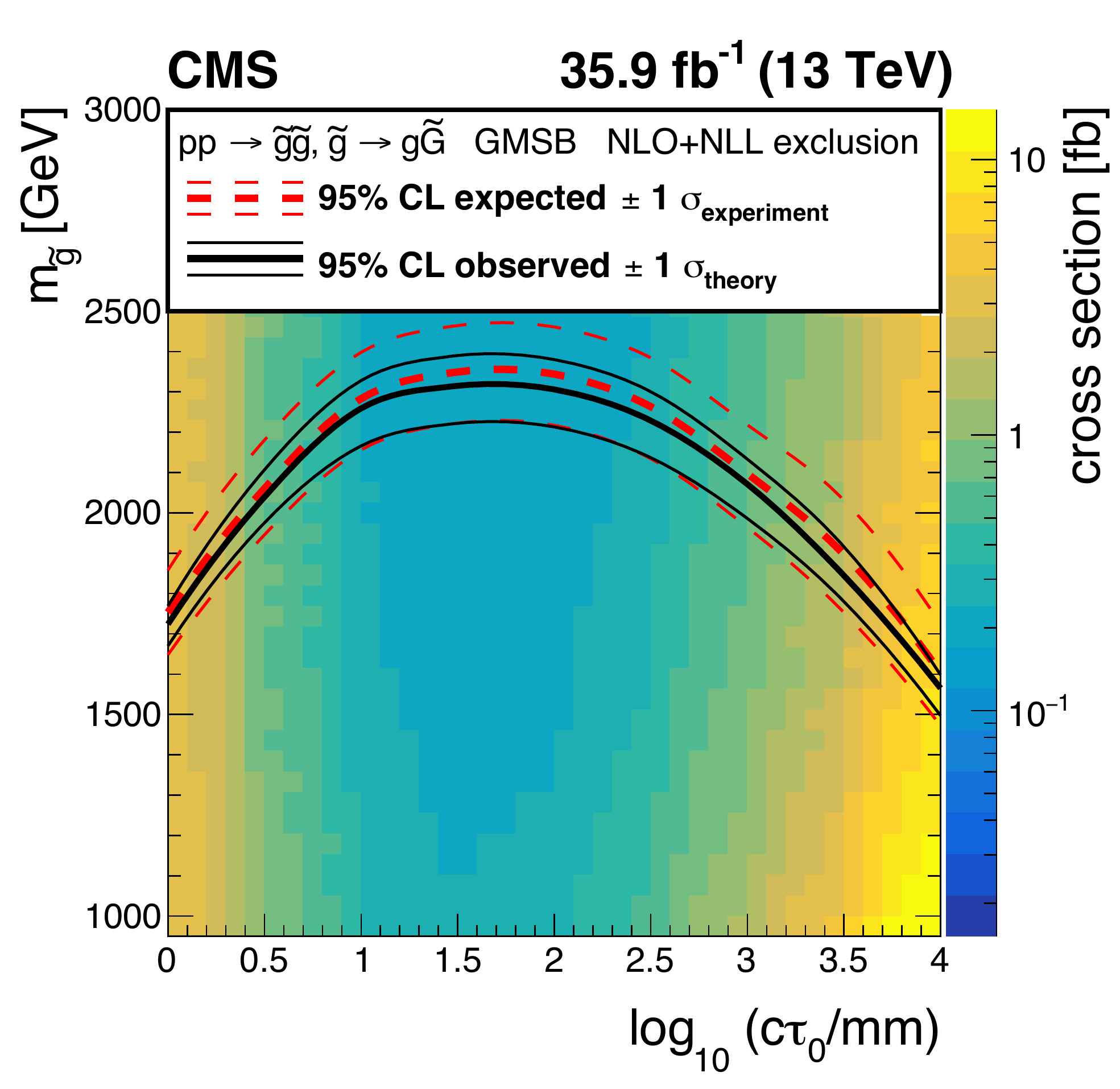}}
\end{minipage}
\caption[]{Likelihood discriminant for data, simulated QCD multijet events, and simulated signal events (left), expected and observed 95\% CL upper limits on the pair production cross section of the long-lived particle X, assuming a 100\% branching fraction for X to decay to a quark-antiquark pair, shown at different X masses and proper decay lengths for the Jet-Jet model  (centre), and expected and observed 95\% CL limits for the long-lived gluino model in the mass-lifetime plane, assuming a 100\% branching fraction for $\tilde g \rightarrow g \tilde G$ decays (right).\cite{CMS-EXO-18-007}}
\label{fig:CMS-EXO-18-007}
\end{figure}

CMS has performed a search for long-lived particles decaying into jets \cite{CMS-EXO-18-007}, testing a GMSB SUSY model with a long-lived gluino decaying to a gluon and a gravitino, and the so-called Jet-Jet model, where long-lived scalar neutral particles X are pair-produced through a $2 \rightarrow 2$ scattering process, mediated by an off-shell Z boson propagator.
Each LLP can have a decay vertex displaced from the production vertex by up to 55 cm in the plane transverse to the LHC beams. The jets were reconstructed from energy deposits in the calorimeter towers. The data sample corresponds to an integrated luminosity of 35.9 fb$^{-1}$, collected with a dedicated displaced jet-trigger, which required $H_\mathrm{T}$ to be larger than 350 GeV, with $H_\mathrm{T}$ defined as the scalar sum of the transverse momenta ($p_\mathrm{T}$) of all jets satisfying $p_\mathrm{T} >$ 40 GeV and $\vert \eta \vert <$ 2.5. The online event selection also required at least two jets with $p_\mathrm{T} >$ 40 GeV within $\vert \eta \vert <$ 2.0, at most two associated prompt tracks with transverse impact parameters with respect to the leading primary vertex smaller than 1 mm, and at least one associated displaced track with a transverse impact parameter larger than 0.5 mm, and an impact parameter significance larger than 5. 
The main background comes from QCD jets, which are suppressed using a likelihood discriminant based on three variables based on track, jet and vertex information. Figure \ref{fig:CMS-EXO-18-007} (left) shows the likelihood discriminants for data, simulated QCD multijet events, and simulated signal events. 
The cutoff value of 0.9993 for the event selection was determined by optimising the signal sensitivity for the Jet-Jet model across proper decay lengths from 1 to 1000~mm and X masses from 100 to 1000~GeV.
Figure \ref{fig:CMS-EXO-18-007} (centre) shows the expected and observed 95\% confidence level (CL) upper limits on the pair production cross section of the long-lived particle X, assuming X to decay exclusively to a quark-antiquark pair, for different particle masses $m_\mathrm{X}$ and proper decay lengths for the Jet-Jet model, while Fig.~\ref{fig:CMS-EXO-18-007} (right) depicts the corresponding limits for the GMSB SUSY model, assuming a 100\% branching fraction for $\tilde g \rightarrow g \tilde G$ decays. The solid (dashed) lines represent the observed (median expected) limits. The shaded bands represent the regions containing 68\% of the distributions of the expected limits under the background-only hypothesis. 
For the Jet-Jet model the limits are most stringent for proper decay lengths between 3 and 100~mm. For smaller decay lengths, the limits become less restrictive because of the vetoes 
on prompt activity. Since the tracking efficiency decreases with larger displacement, the limits also become less stringent for larger decay lengths. For the GMSB model gluino masses up to 2300~GeV are excluded for proper decay lengths between 20 and 110~mm.

\subsection{Delayed jets}\label{subsec:DelayedJets}
CMS has also searched for delayed jets resulting from displaced decays of long-lived particles in the electromagnetic calorimeter (ECAL) \cite{CMS-EXO-19-001}. The delay is expected to be a few nanoseconds for a TeV scale particle which travels a distance of the order of 1~m before decaying. Events with such particles would escape reconstruction in a tracker-based search because of the inability to reconstruct tracks for decay points separated from the primary vertex by more than about 50~cm in the plane perpendicular to the beam axis.
A similar GMSB benchmark model as in the previous subsection was used as a benchmark. This model contains pairs of long-lived gluinos that each decay into a gluon, which forms a jet, and a gravitino, which escapes the detector causing significant missing energy in the event. The analysis exploits the full Run 3 dataset of 137 fb$^{-1}$. Its specificity is the use of timing information, which leads to a significant gain in sensitivity \cite{Liu}. It is provided by the lead tungstate crystal calorimeter in the barrel detector region, read out by silicon avalanche photodiodes (APD) and yielding a time resolution of 200~ps. 
The online trigger required missing energy larger than 120~GeV. 
Particle flow was not used for the delayed jet reconstruction due to the non-standard tracker component, only calorimeter clustering. In addition to missing energy, the signal selection required jets to have a delay $t_{\mathrm{jet}} > 3$~ns. Crystal cells within a distance $\Delta R < 0.4$ with respect to the jet axis have been taken into account, where $t_{\mathrm{jet}}$ is given by the median cell time, and $t = 0$~ns is defined as the arrival time of a particle travelling from the origin to a cell with velocity $c$. 

\begin{figure} [hbt]
\begin{minipage}{0.50\linewidth}
\centerline{\includegraphics[width=0.9\linewidth]{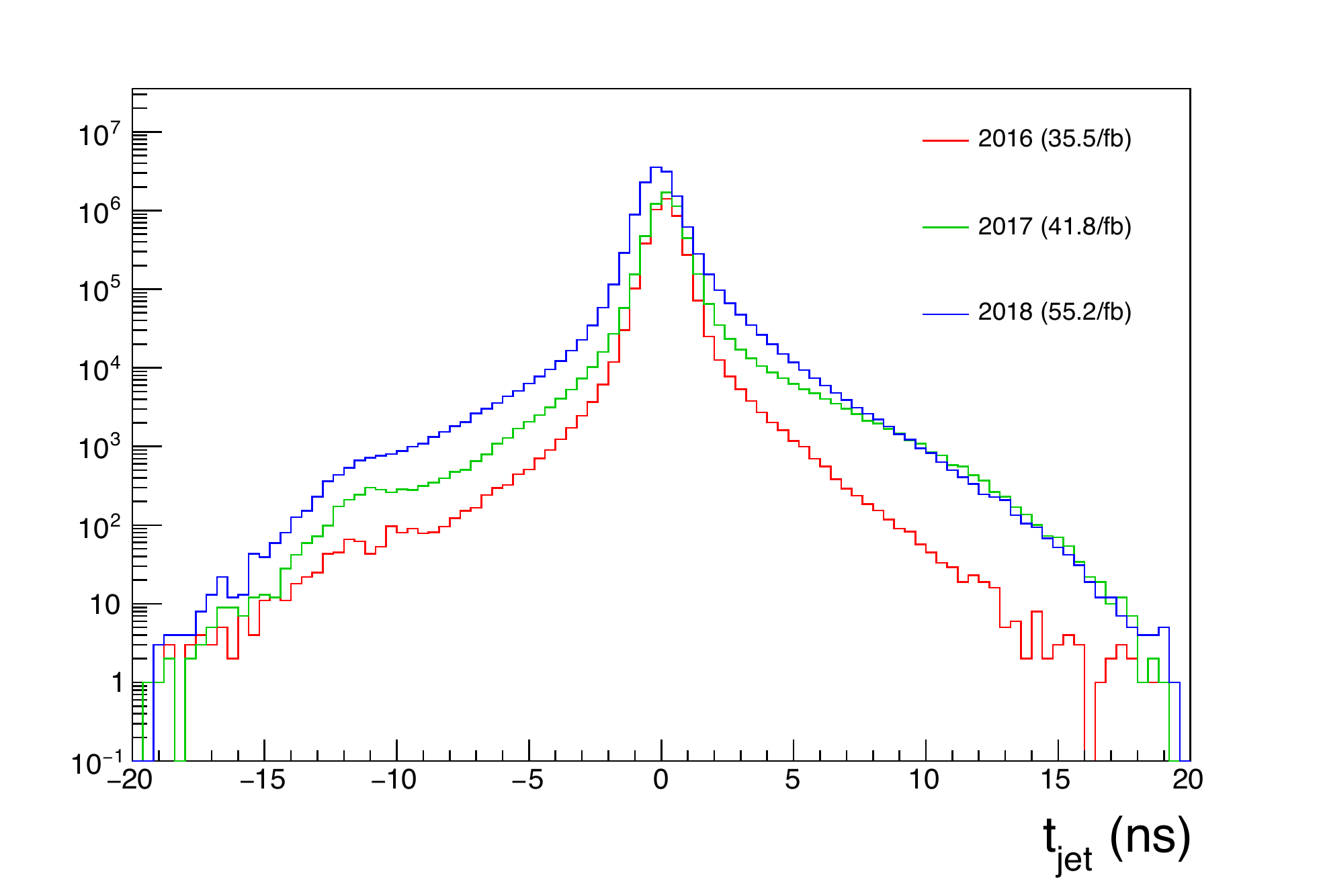}}
\end{minipage}
\hfill
\begin{minipage}{0.50\linewidth}
\centerline{\includegraphics[width=0.9\linewidth]{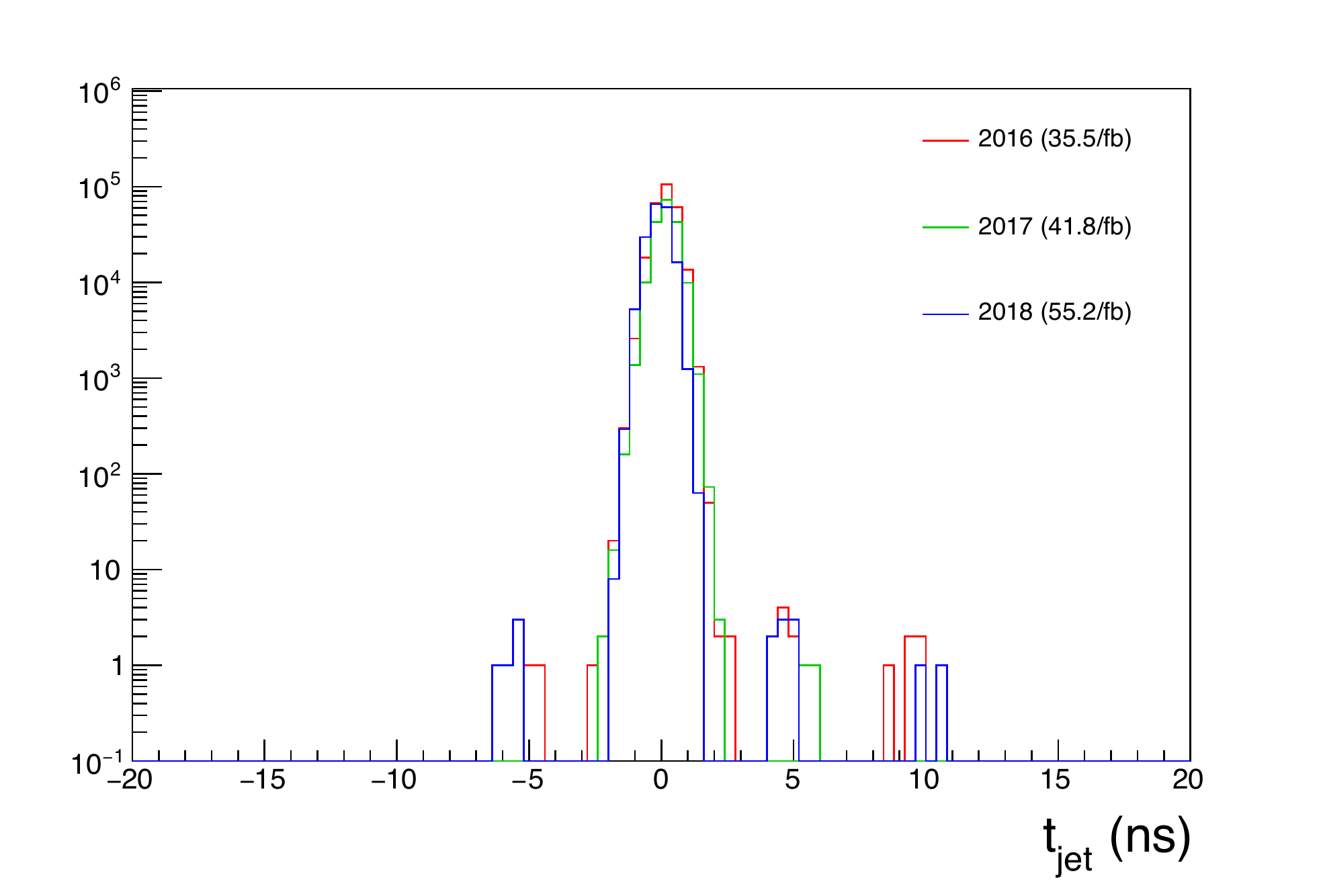}}
\end{minipage}
\caption[]{Jet timing distributions for 2016, 2017 and 2018 data, before (left), and after jet cleaning (right).\cite{CMS-EXO-19-001}}
\label{fig:jetcleaning}
\end{figure}

It was possible to perform the analysis of the full LHC Run 2 dataset at this time, because no Monte Carlo simulations of the backgrounds had been necessary. They were estimated solely from control regions with data. Backgrounds include ECAL resolution tails, direct APD hits, in-time and out-of-time pileup, cosmic muon deposits in the electromagnetic calorimeter, beam halo and satellite bunches. The latter occur in steps of 2.5~ns. The main backgrounds are primarily reduced by jet cleaning, applying several requirements. The number of ECAL cells in a jet must exceed 25, to reject noise and direct APD hits. The hadronic energy fraction, defined as $E_\mathrm{ECAL}/(E_\mathrm{ECAL}+E_\mathrm{HCAL}$), has to be larger than 0.2. The fraction of tracks associated to the primary vertex must be less than 1/12, to reject jets with a mismeasured time or originating from satellite bunch collisions.
The spread in the time of the constituent cells of a signal jet is required to be small, since all the component cells originate from the same delayed jet. The good efficiency of the jet cleaning procedure can be seen from Fig.~\ref{fig:jetcleaning}.

\begin{figure} [hbt]
\begin{minipage}{0.52\linewidth}
\centerline{\includegraphics[width=0.9\linewidth]{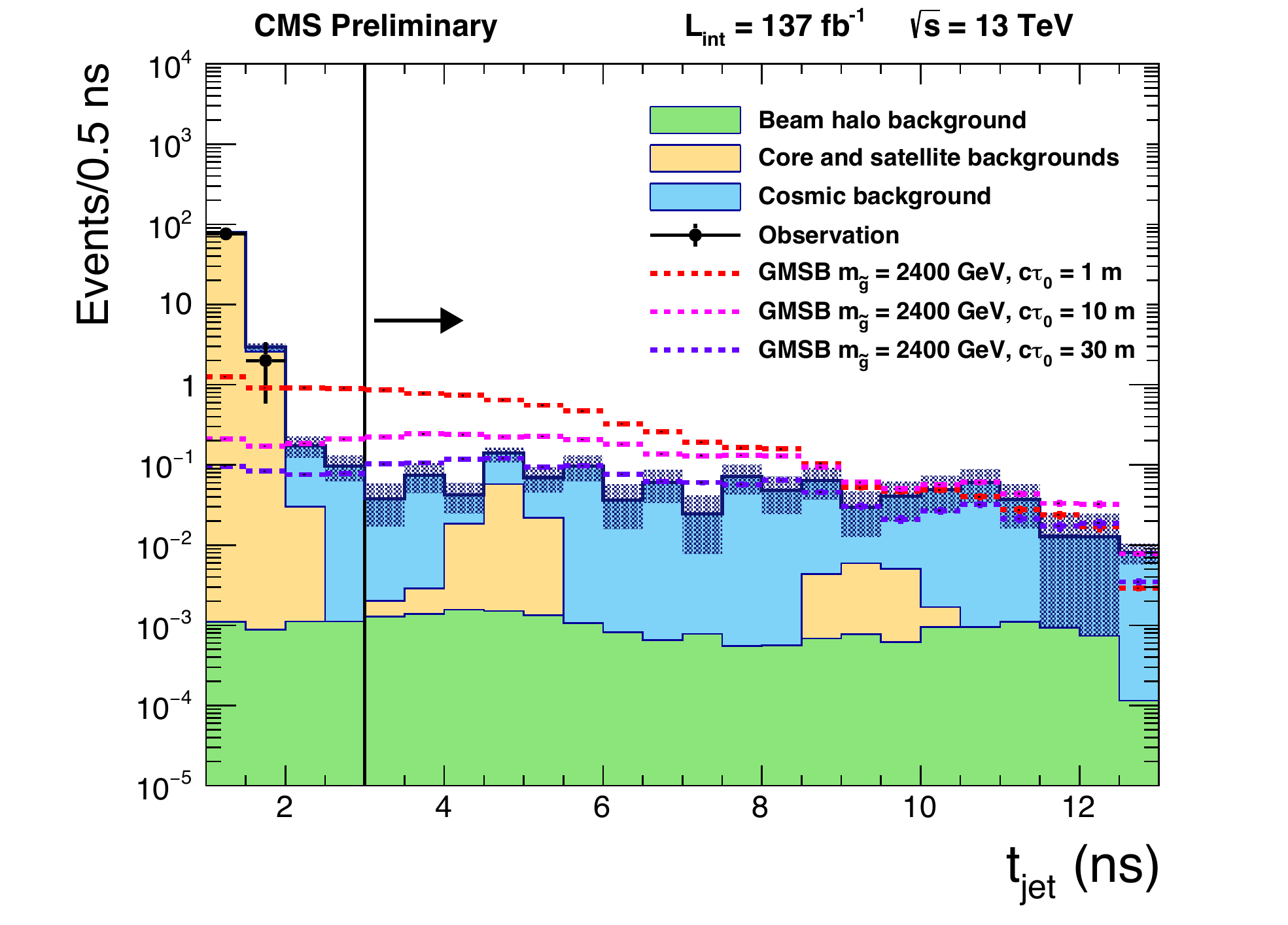}}
\end{minipage}
\hfill
\begin{minipage}{0.38\linewidth}
\centerline{\includegraphics[width=0.9\linewidth]{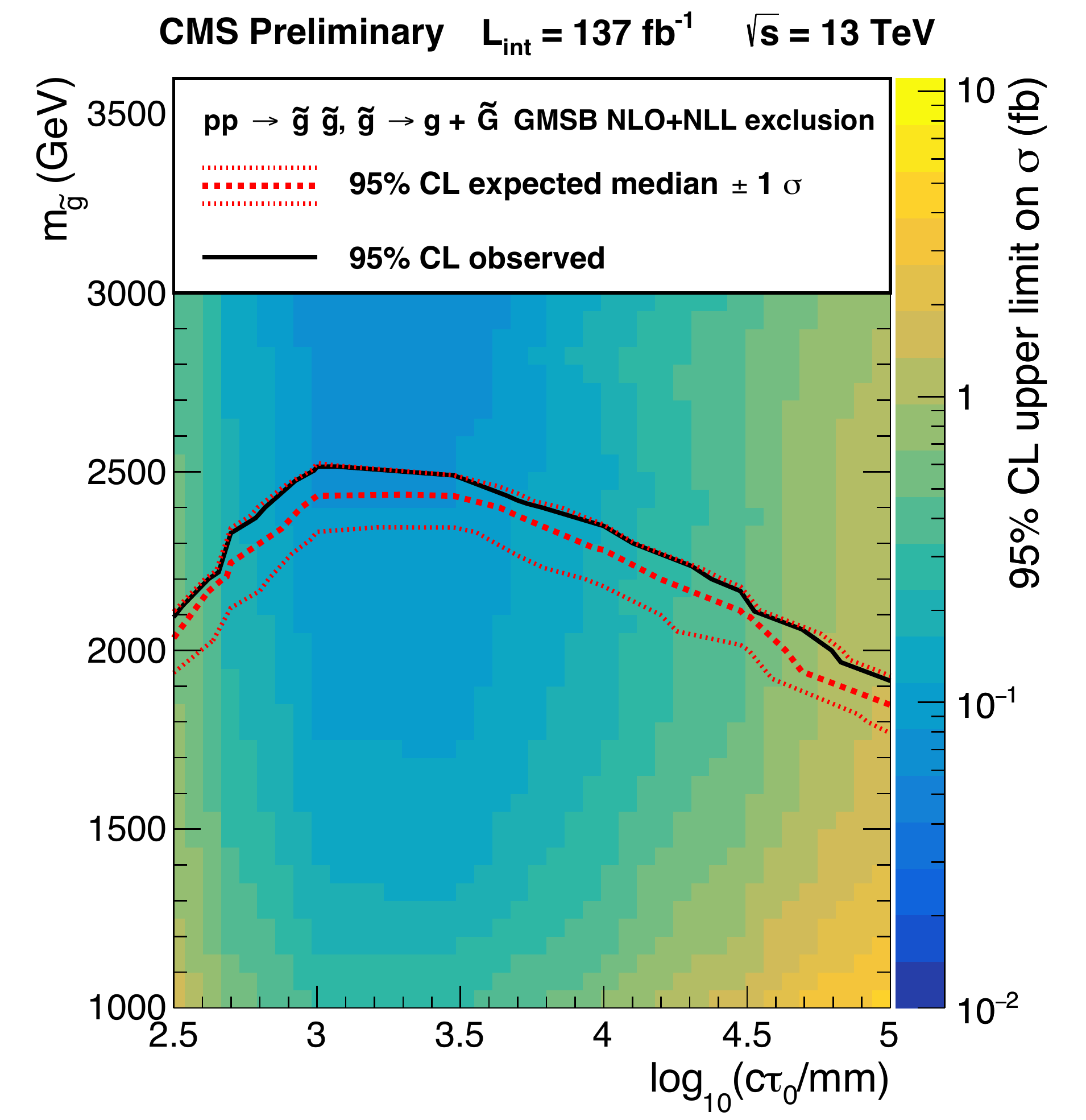}}
\end{minipage}
\caption[]{Timing distribution of the backgrounds in the signal region, compared to predictions for the GSMB model with representative parameters (left); 95\% CL observed upper limits on $\sigma$ in the gluino mass and $c\tau_0$ plane after all selections (right).\cite{CMS-EXO-19-001} The area enclosed by the thick black curve represents the observed exclusion region, while the dashed red lines indicate the expected limits and their $\pm 1$ standard deviation ranges. The thin black lines show the effect of the theoretical uncertainties on the signal cross section. }
\label{fig:EXO-19-001-results}
\end{figure}

The jet timing distribution of the backgrounds in the signal region, compared to signals predicted by the above-mentioned GMSB model for different gluino masses and lifetimes is depicted in Fig.~\ref{fig:EXO-19-001-results} (left).
The obtained results in Fig.~\ref{fig:EXO-19-001-results} (right) show that the signal reach can be significantly extended compared to tracker-based searches by using timing information of electromagnetic energy deposits \cite{PhysRevD.97.052012}$^,$ \cite{PhysRevD.99.032011}$^,$ \cite{PhysRevD.98.092011}. The interpretation within the model yields exclusions of gluino masses up to 2500 GeV (2150 GeV) for proper lifetimes of 1m (30 m).  

\subsection{Trackless jets}\label{subsec:TracklessJets}
\begin{figure} [hbt]
\begin{minipage}{0.34\linewidth}
\centerline{\includegraphics[width=0.9\linewidth]{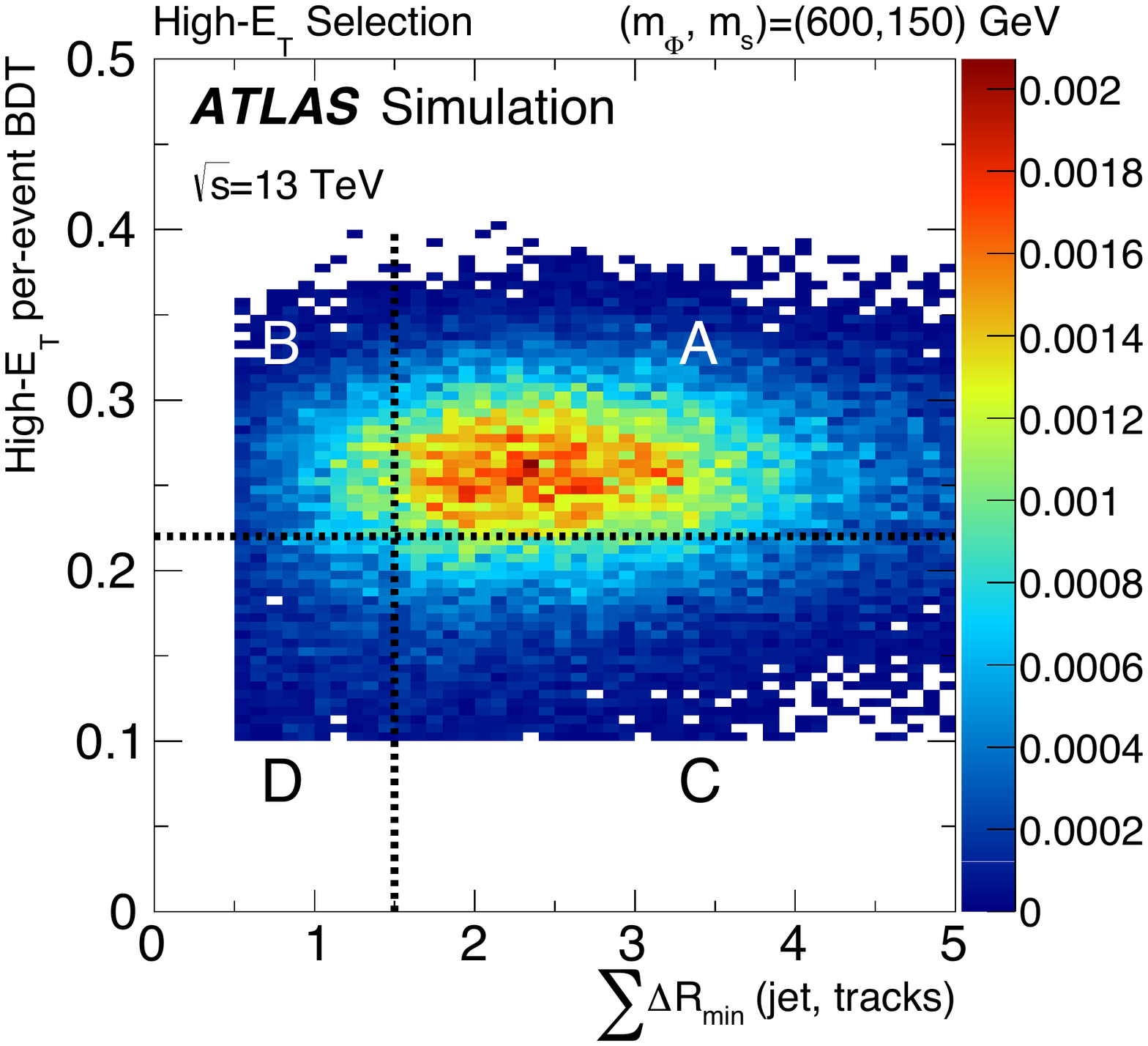}}
\end{minipage}
\hfill
\begin{minipage}{0.32\linewidth}
\centerline{\includegraphics[width=0.9\linewidth]{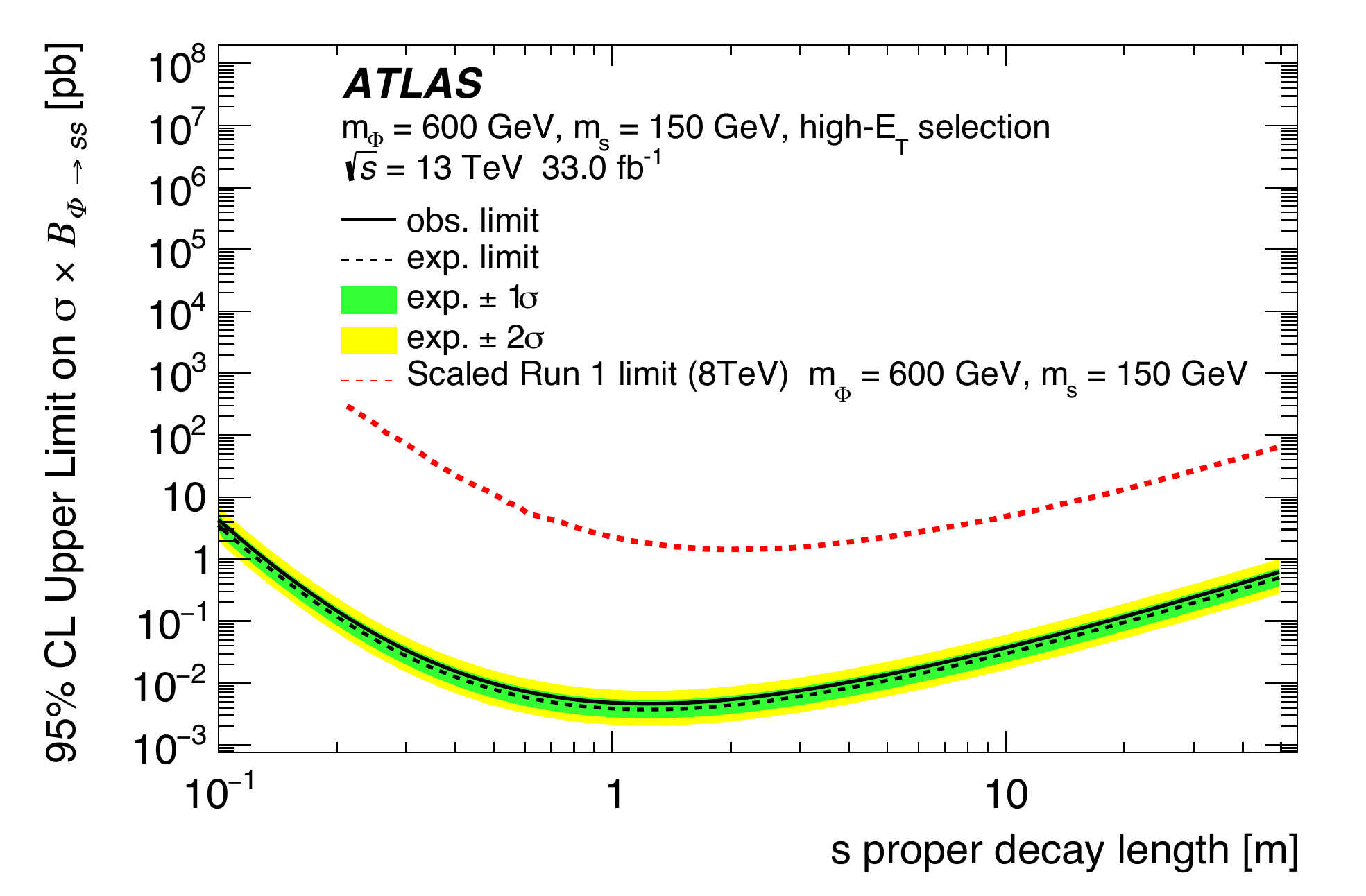}}
\end{minipage}
\hfill
\begin{minipage}{0.32\linewidth}
\centerline{\includegraphics[width=0.9\linewidth]{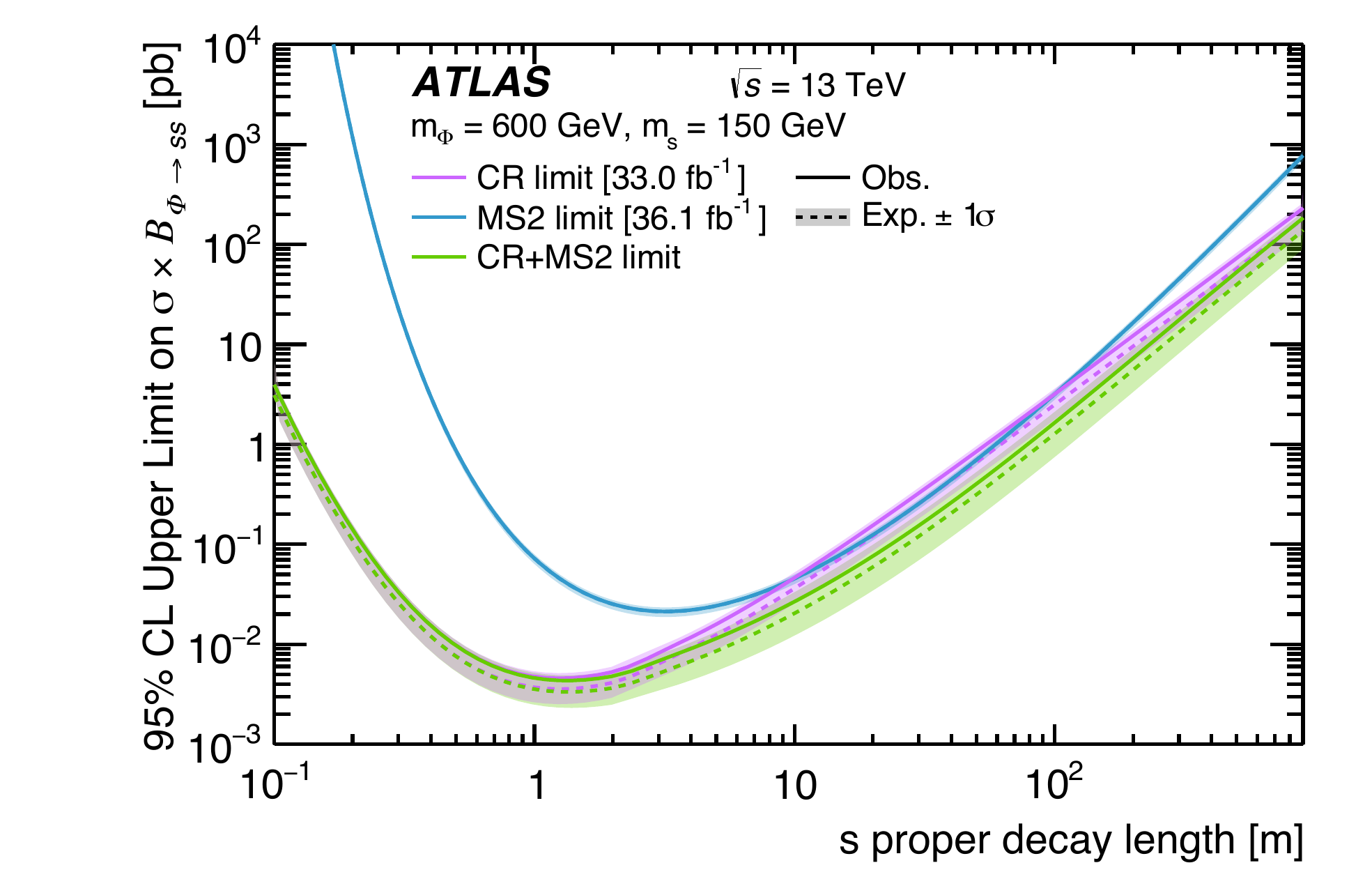}}
\end{minipage}
\caption[]{$\Sigma \Delta R_\mathrm{min}$(jet, tracks) versus per-event BDT for a simulated signal sample after event selection for the high-$E_\mathrm{T}$ sample (left); observed cross section times branching fraction limits for a mediator mass of $m_\Phi = 600$~GeV (centre); combined results for CalRatio and muon system analyses for $m_\Phi = 600$~GeV (right).\cite{ATLAS_LLtracklessJets}}
\label{fig:ATLAS_LLtracklessJets}
\end{figure}

A search for pairs of neutral, long-lived scalar particles decaying to trackless jets in the calorimeter has been performed by the ATLAS Collaboration \cite{ATLAS_LLtracklessJets}. The signature consists of displaced jets in the hadron calorimeter of the outer edge of the electromagnetic calorimeter. The analysis probes proper decay lengths between a few centimetres and a few tens of metres. As a benchmark, a simplified model with a hidden sector that is connected to the standard model  through a heavy neutral boson $\Phi$. It decays to two scalars $s$, which in turn  decay to fermion-antifermion pairs. The data consisted of two samples of 10.8~fb$^{-1}$ and 33.0~fb$^{-1}$, recorded with a low- and a high-$E_\mathrm{T}$ trigger, respectively. The event selection required at least two trackless jets with a low fraction of electromagnetic energy, also called CalRatio (CR) jets. To select events with trackless jets, an event-level variable, $\Delta R_\mathrm{min}$(jet, tracks), is used, where $\Delta R_\mathrm{min}$(jet, tracks) is defined as the angular distance between the jet axis and the closest track with $p_\mathrm{T} > 2$~GeV, and $\Sigma \Delta R_\mathrm{min}$(jet, tracks) is calculated by summing this distance over all the clean jets with $p_\mathrm{T} > 50$~GeV. Machine learning techniques have been used for the analysis, namely a neural network to determine the jet origins, and a boosted decision tree (BDT) classifier for jets. The backgrounds, estimated with an ABCD method, consist mainly of QCD multijets and beam-induced backgrounds. Fig.~\ref{fig:ATLAS_LLtracklessJets} (left) demonstrates the efficiency of the selection. 
Limits at 95\%~CL have been set on the cross section times branching fraction $\sigma(\Phi) \times \mathcal{B} ( \Phi \rightarrow s\bar s)$. For mediator masses $m_\Phi$ of 400, 600, and  1000~GeV values above 0.1~pb are excluded between 12~cm and 9~m, 7~cm and 20~m, and 4~cm and 35~m respectively, depending on $m_s$. The limit curve for $m_\Phi = 600$~GeV and $m_s = 150$~GeV is shown in Fig.~\ref{fig:ATLAS_LLtracklessJets} (centre). The corresponding Run~1 curve \cite{Run1LL} is scaled by the ratio of parton luminosities for gluon-gluon fusion between 13 and 8~TeV. The right-hand figure shows results for this CalRatio jet analysis as well as a combination with a displaced jet analysis in the muon system (MS) \cite{Aaboud:2018aqj}. 
For masses $m_\Phi \geq 600$~GeV, the CR analysis is in general more sensitive than the MS analysis. The combination provides a modest improvement on the CR-only limit at long decay lengths.

\subsection{Emerging jets}\label{subsec:EmergingJets}
\begin{figure} [hbt]
\begin{minipage}{0.33\linewidth}
\centerline{\includegraphics[width=0.9\linewidth]{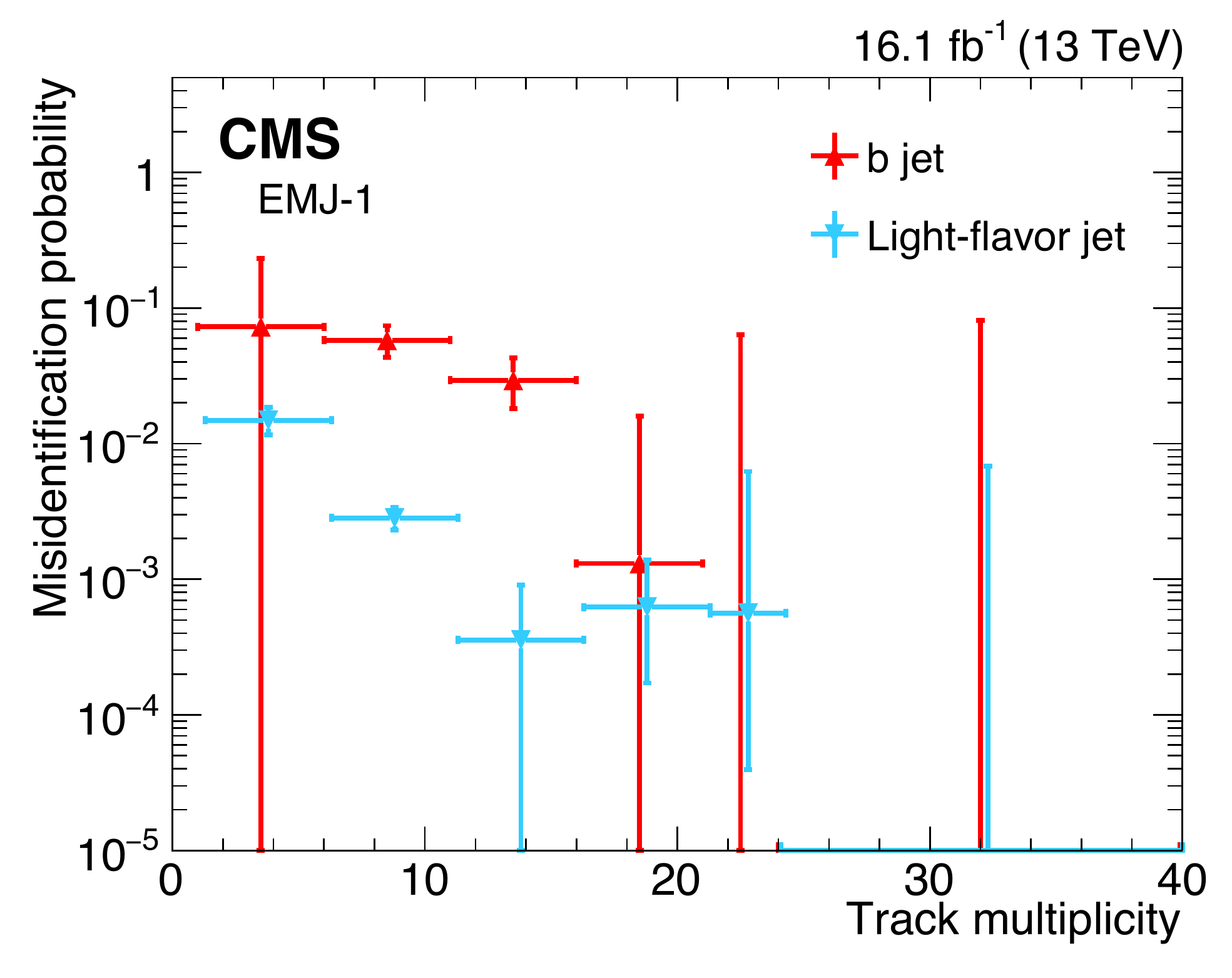}}
\end{minipage}
\hfill
\begin{minipage}{0.32\linewidth}
\centerline{\includegraphics[width=0.9\linewidth]{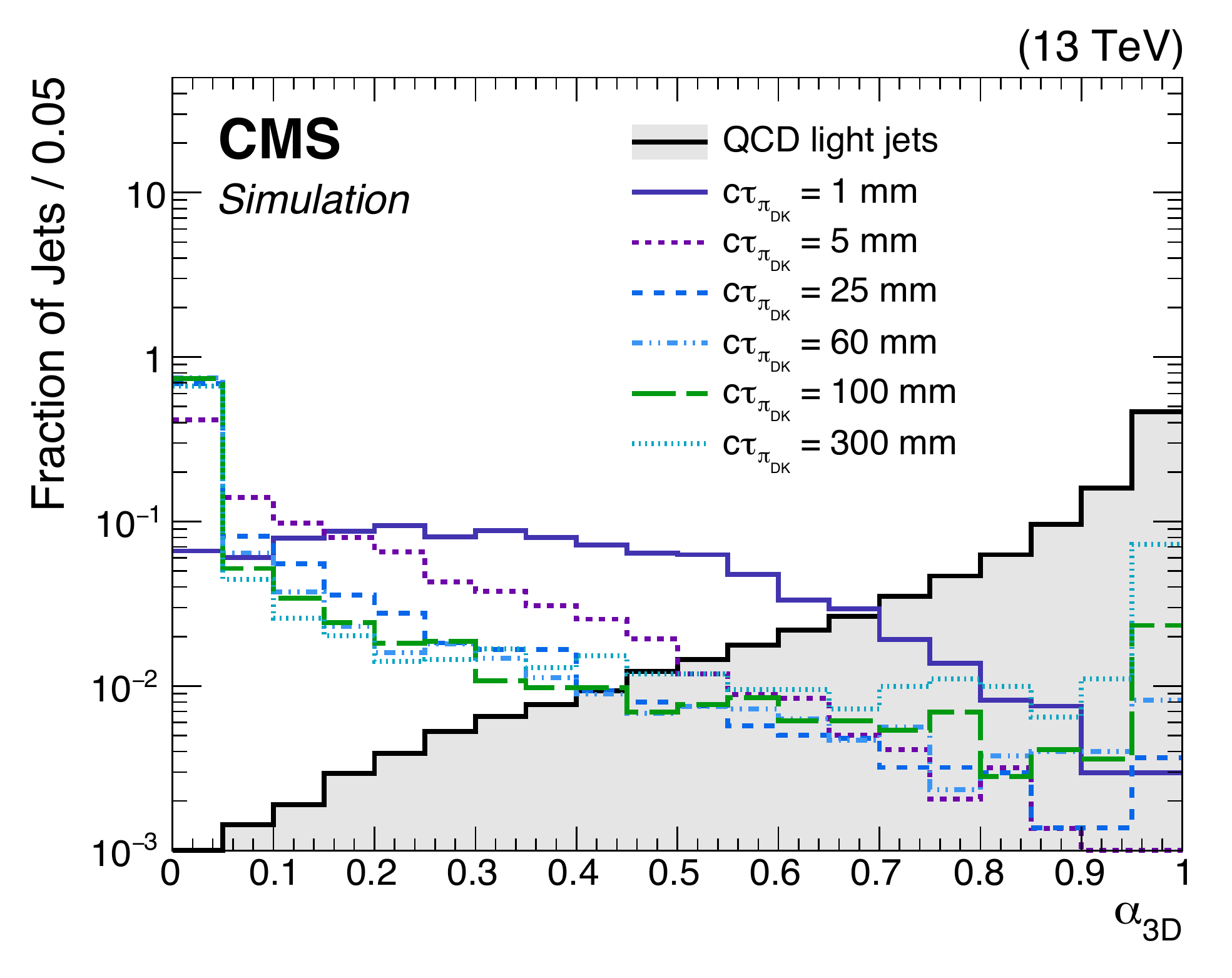}}
\end{minipage}
\hfill
\begin{minipage}{0.32\linewidth}
\centerline{\includegraphics[width=0.9\linewidth]{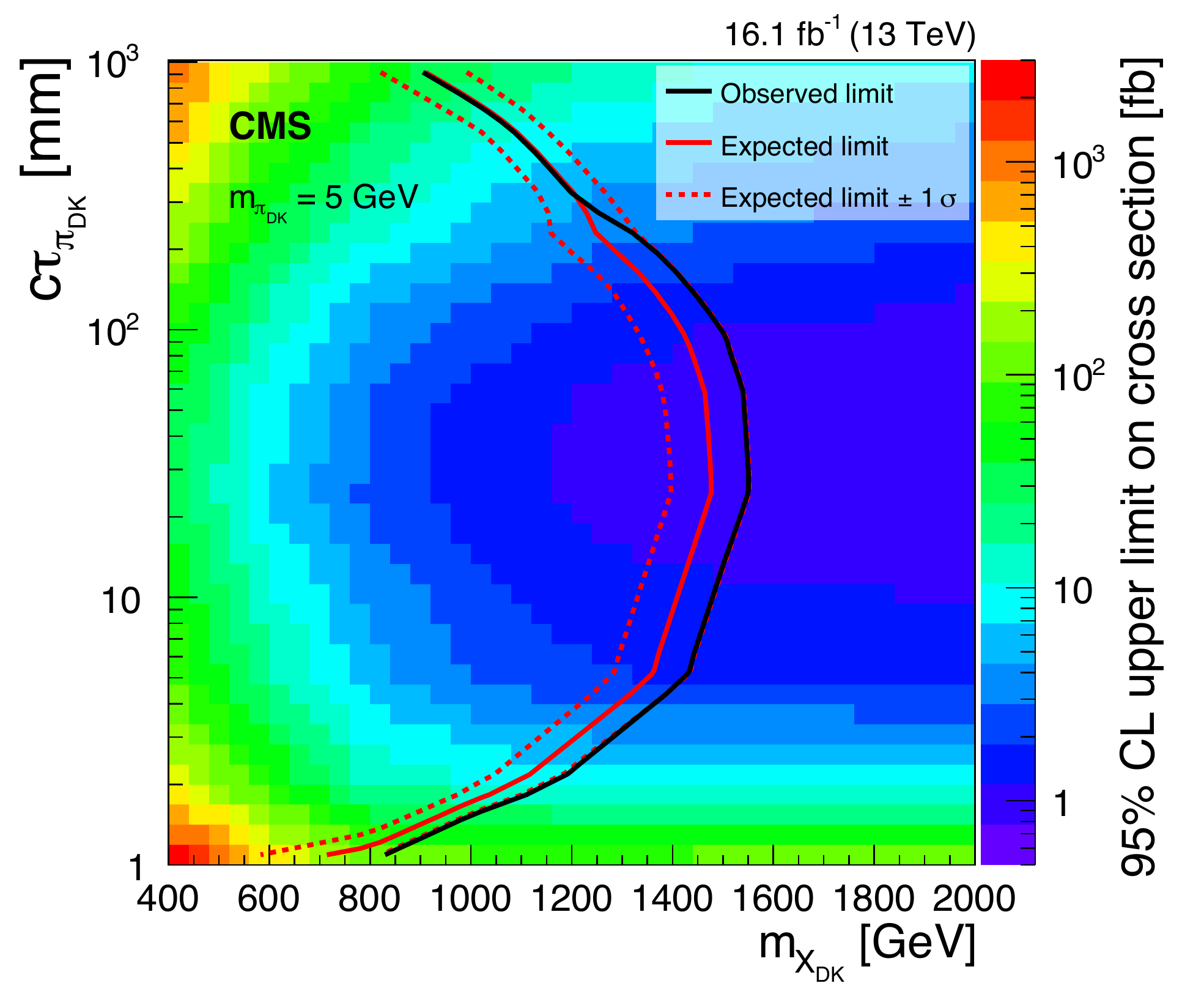}}
\end{minipage}
\caption[]{Misidentification rate as a function of track multiplicity for b jets and light-quark jets (left); $\alpha_{3\mathrm{D}}$ for background (black) and for signals with a dark mediator mass of 1~TeV and a dark pion mass of 5 GeV for dark pion proper decay lengths ranging from 1 to 300~mm (centre); upper limits at 95\% CL on the signal cross section and signal exclusion contours derived from theoretical cross sections for a model with a dark pion mass of 5 GeV (right).\cite{CMSemergingjets}}
\label{fig:emergingjets}
\end{figure}

Emerging jets are characteristic for models with a composite dark sector. CMS has studied such a model \cite{Schwaller}, which contains a heavy mediator $\mathrm{X_{DK}}$ that couples both to dark sector and SM particles. The experimental signature consists of four high-$p_\mathrm{T}$ jets, two from down quarks and two from dark quarks \cite{CMSemergingjets}.  The dark quark jets contain many displaced vertices arising from the decays of dark pions ($\pi_{\mathrm{DK}}$) produced in the dark parton shower and fragmentation.The trigger was a simple $H_\mathrm{T}$ trigger with a threshold of 900~GeV, and the event selection required two jets with displaced tracks and many different vertices within the jet cone, and two regular jets. The main background is SM four-jet production, where jets are tagged as emerging either because they contain long-lived B mesons or reflect track misreconstruction. It is estimated with a data-driven method and depends on parton flavour and track multiplicity, as can be seen from Fig.~\ref{fig:emergingjets} (left). 
The central figure shows the variable $\alpha_{3\mathrm{D}}$, which quantifies the fraction of jet $p_\mathrm{T}$ associated with prompt tracks, for ordinary QCD jets and for signal jets arising from dark pions of mass 5~GeV with different lifetimes. Figure~ref{fig:emergingjets} (right) shows upper limits at 95\% CL on the signal cross section and signal exclusion contours derived from theoretical cross sections for a model with a dark pion mass of 5 GeV in the plane of the mediator mass and the dark pion lifetime. The solid red contour is the expected upper limit, with its one standard- deviation region enclosed in red dashed lines. The solid black contour is the observed upper limit. The region to the left of the observed contour is excluded. Mediator masses between 400 and 1250~GeV are thus excluded for decay lengths 5 to 225~mm.

\subsection{Displaced dimuons}\label{subsec:Displaced Dimuons}
\begin{figure} [hbt]
\vspace*{-1cm}
\begin{minipage}{0.24\linewidth}
\centerline{\includegraphics[width=0.9\linewidth]{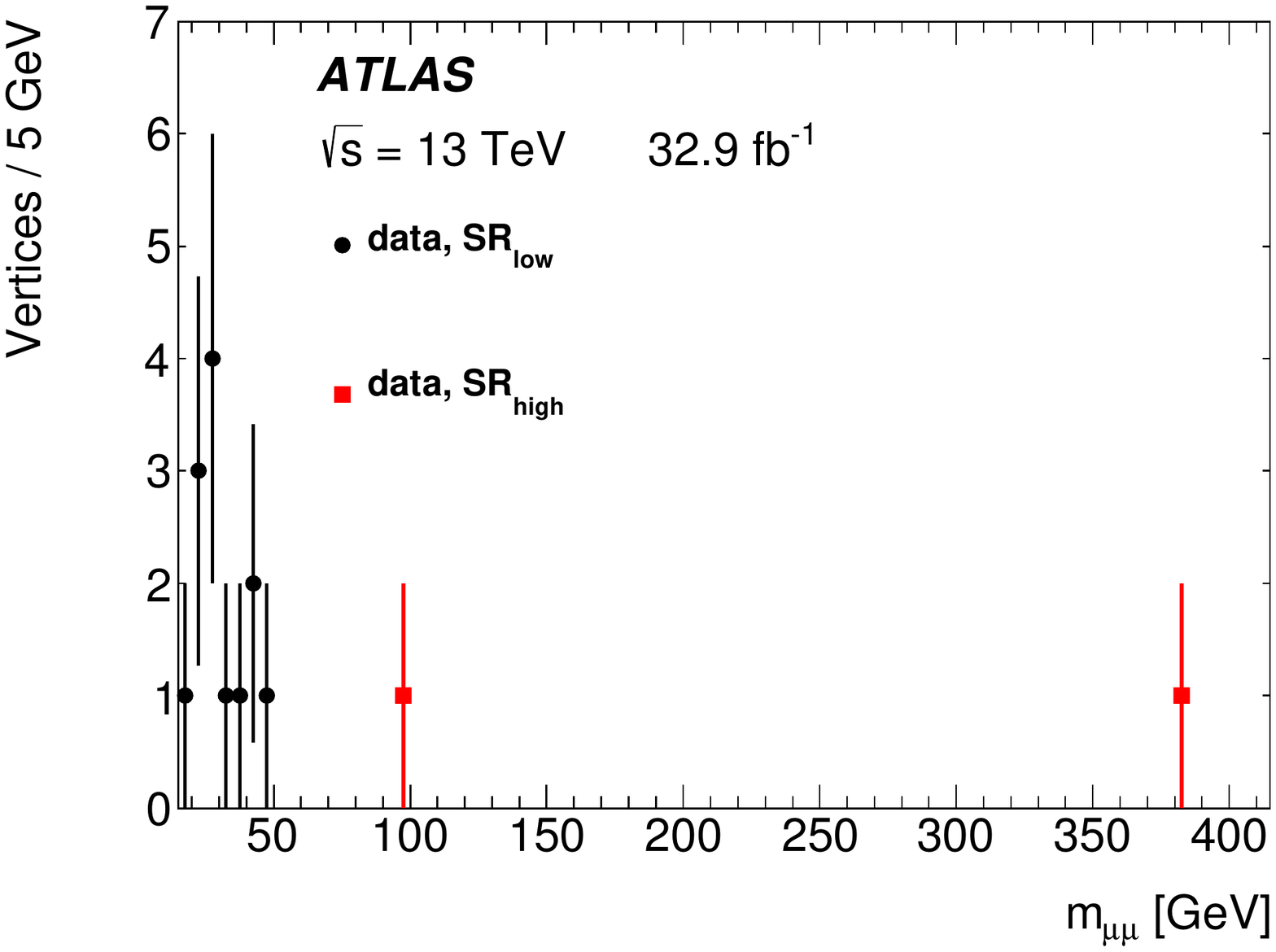}}
\end{minipage}
\hfill
\begin{minipage}{0.24\linewidth}
\centerline{\includegraphics[width=0.9\linewidth]{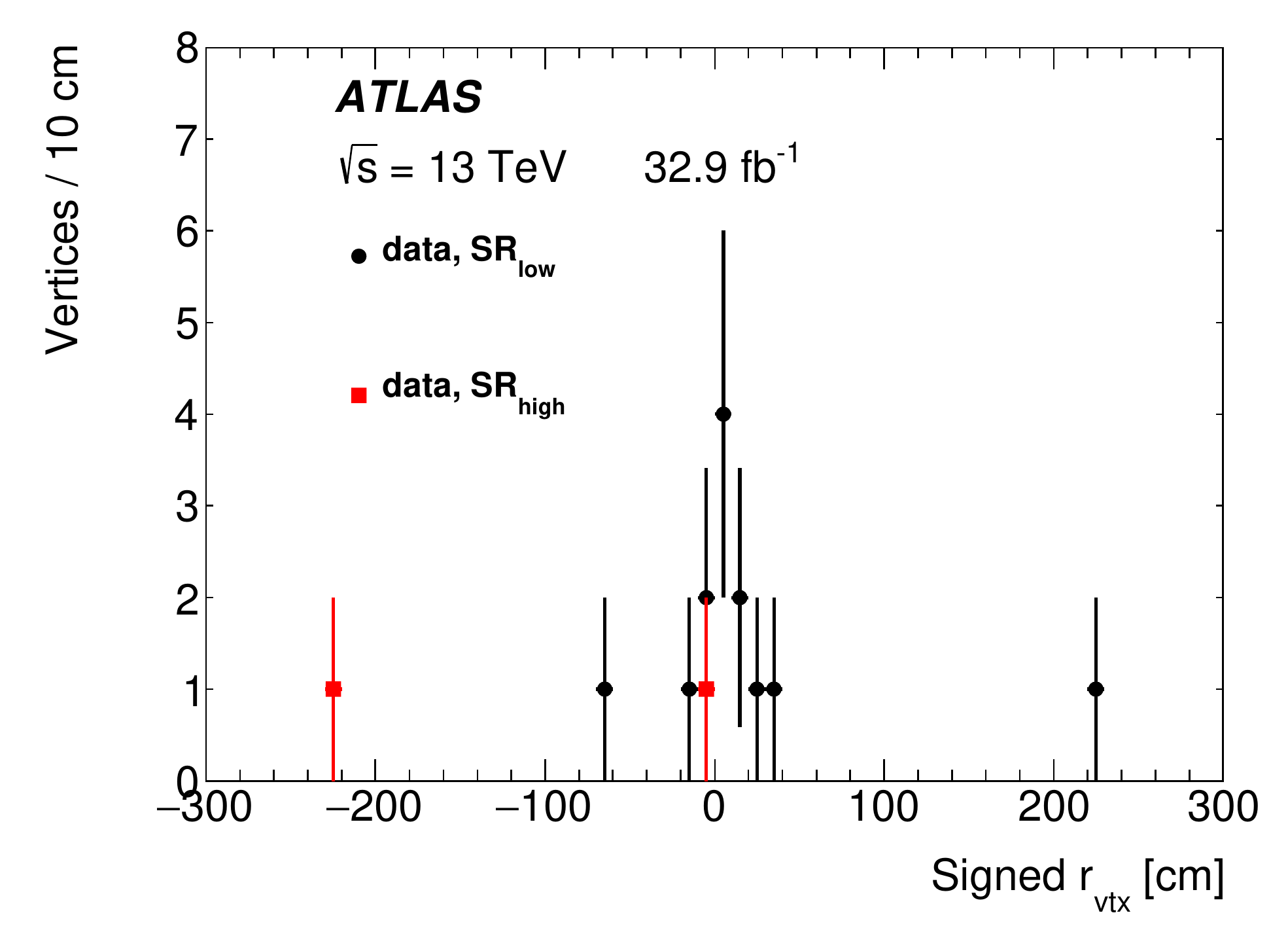}}
\end{minipage}
\begin{minipage}{0.24\linewidth}
\centerline{\includegraphics[width=0.9\linewidth]{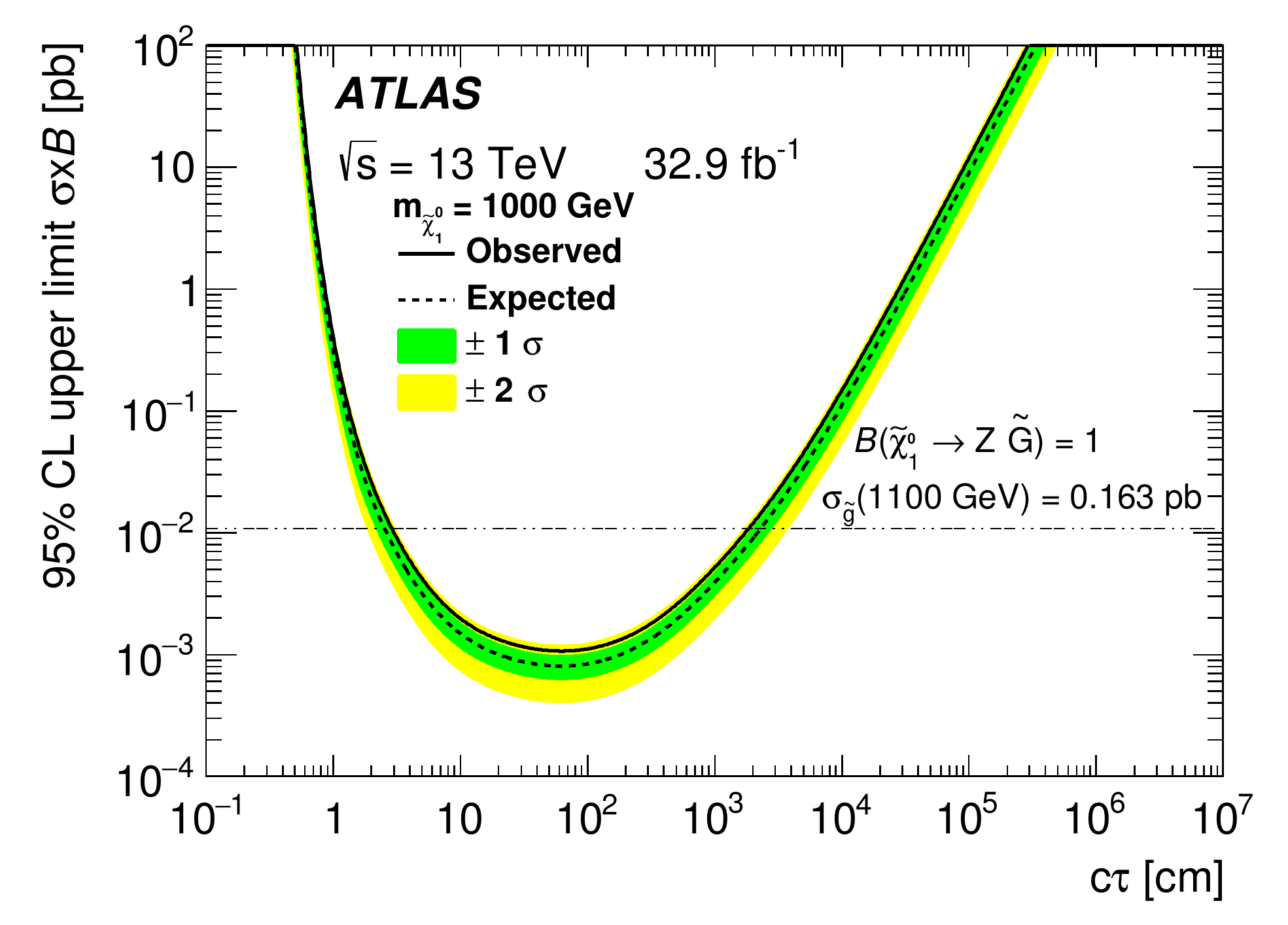}}
\end{minipage}
\hfill
\begin{minipage}{0.24\linewidth}
\centerline{\includegraphics[width=0.9\linewidth]{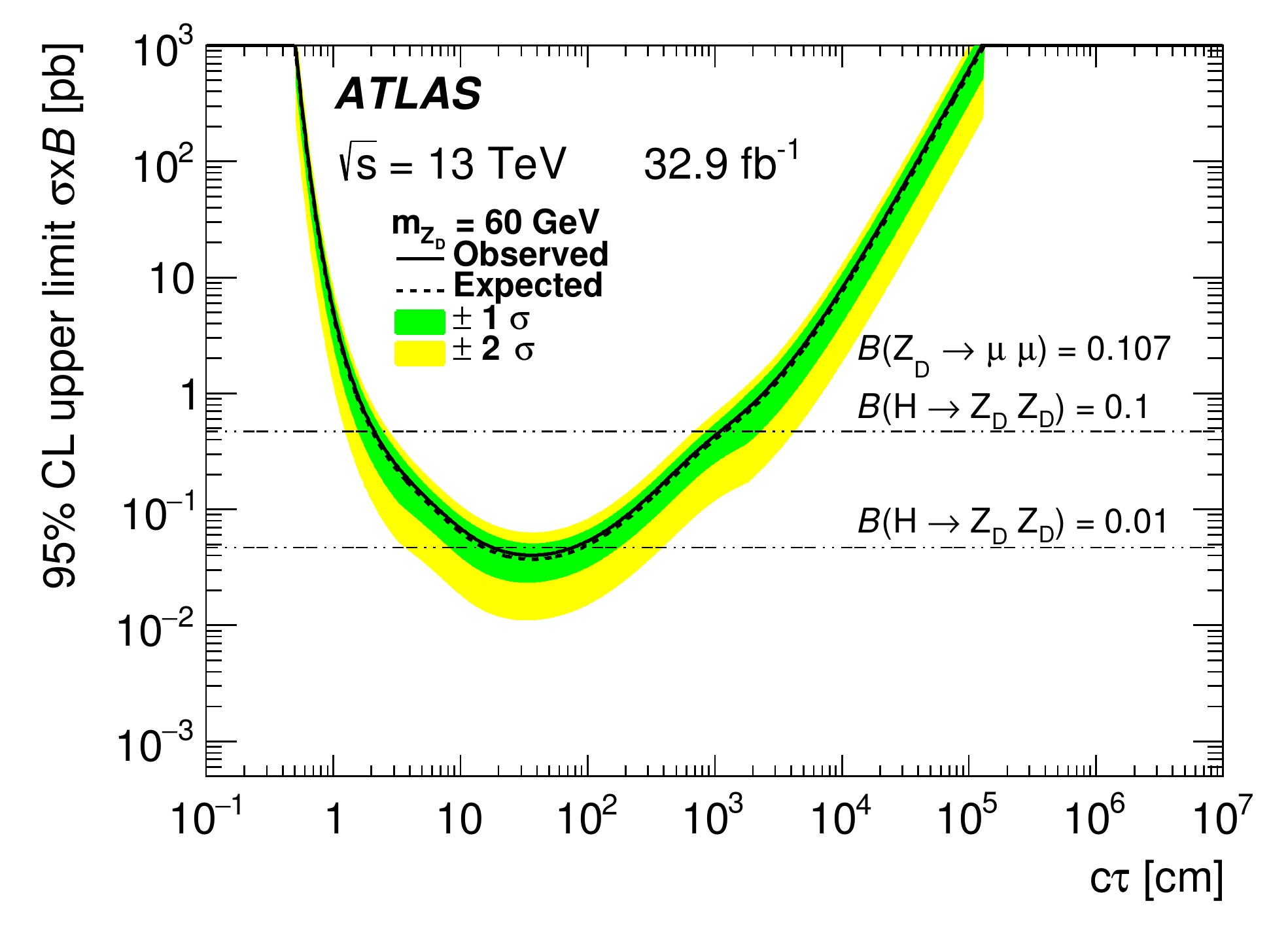}}
\end{minipage}
\vspace*{-1cm}
\caption[]{Dimuon mass and distance of dimuon vertex from interaction point (left); exclusion contours for the GMSB model and the dark photon model (right).\cite{ATLASdimuons}}
\label{fig:dimuons}
\end{figure}

A search for long-lived neutralinos and long-lived dark photons Z$_\mathrm{D}$ has been studied by ATLAS \cite{ATLASdimuons}, using displaced muons as the signature. In the first model the neutralino of a GMSB scenario decays into a gravitino and a Z boson, which in turn decays to dimuons. In the second model two Z$_\mathrm{D}$ arise from the decay of a Higgs boson, with at least one of them decaying to two displaced muons. The trigger was as inclusive as possible, requiring two opposite-sign muons with a vertex displacement of up to 4~m from the interaction point. Both a high-mass and a low-mass trigger were used, with the first one targeting the decay Z$\rightarrow \mu^+\mu^-$, and the second one Z$_\mathrm{D} \rightarrow \mu^+\mu^-$. The high-mass trigger required missing energy and a single muon with $p_\mathrm{T} > 60$~GeV, while the low-mass trigger required collimated dimuons or trimuons with lower transverse momenta, between 20 and 6~GeV. Requiring missing energy compensates the loss in single muon trigger efficiency, which drops from 70\% at the interaction point to 10\% at a distance of 4~m. Backgrounds come mainly from  Drell-Yan, $t \bar t$, single top and multijet production. The dimuon mass and vertex distance distributions for the two trigger samples are shown in Fig.~\ref{fig:dimuons} (left). The events with the highest masses have been identified as a Z and a cosmic ray particle, respectively. No signal has thus been found. Exclusion contours for the two models in the cross section times branching fraction and lifetime plane are depicted in Fig.~\ref{fig:dimuons} (right). Lower and upper lifetime limits of 0.3 and 2400~cm, depending on the model parameters, have been set. 

\section{Conclusions}
No signals for long-lived neutral particles have yet been found in the analysis of the LHC Run~2 data analysed so far by the ATLAS and CMS collaborations. Non-standard trigger, data acquisition and analysis strategies and techniques are being further developed, and results for the full dataset will be published soon.

\end{document}